\documentclass[sn-standardnature,iicol]{sn-jnl}
\usepackage{natbib}
\usepackage{siunitx}
\usepackage{amssymb}
\usepackage{times}
\usepackage{gensymb}
\usepackage{booktabs}
\usepackage{astjnlabbrev}

\usepackage{amsmath, array, multirow, rotating, makecell}

\usepackage[version=4]{mhchem}

\newcommand\arcsec{''}
\newcommand\micron{$\mu$m}
\newcommand\microns{$\mu$m}

\begin{document}

\title[]{\centering Horizontal transport as a source of disequilibrium chemistry on the nightside of a hot exoplanet}
%=============================================================%%

\author*[1]{\fnm{Vivien} \sur{Parmentier}}\email{vivien.parmentier@oca.eu}
\author[2]{\fnm{Kevin B.} \sur{Stevenson}}
\author[3]{\fnm{Luis} \sur{Welbanks}}

\author[4]{\fnm{Jake} \sur{Taylor}}
\author[5]{\fnm{Everett} \sur{Schlawin}}
\author[6]{\fnm{Louis-Philippe} \sur{Coulombe}}
\author[7]{\fnm{Yao} \sur{Tang}}
\author[3]{\fnm{Mike} \sur{Line}}
\author[8]{\fnm{Hinna} \sur{Shivkumar}}
\author[9]{\fnm{Xianyu} \sur{Tan}}

%Alphabetical order

\author[10]{\fnm{Jacob L.} \sur{Bean}}
\author[11,8,15]{\fnm{Jean-Michel} \sur{D\'esert}}
\author[7]{\fnm{Jonathan  J.} \sur{Fortney}}
\author[12]{\fnm{Peter} \sur{Gao}}
\author[13]{\fnm{Mark} \sur{Hammond}}
\author[10,14]{\fnm{Eliza M.-R.} \sur{Kempton}}
\author[13]{\fnm{Thaddeus D.} \sur{Komacek}}
\author[14]{\fnm{Megan} \sur{Weiner Mansfield}}

\affil*[1]{\orgdiv{Laboratoire Lagrange}, \orgname{Université de la Côte d’Azur, Observatoire de la Côte d’Azur, CNRS}, \orgaddress{\street{Bd de l’Observatoire}, \city{Nice}, \postcode{06304}, \country{France}}}
\affil[2]{\orgdiv{Johns Hopkins}, \orgname{Applied Physics Laboratory}, \city{Laurel}, \postcode{MD }, \country{USA}}
\affil[3]{\orgdiv{School of Earth and Space Exploration}, \orgname{Arizona
State University}, \city{Tempe}, \state{AZ}, \country{USA}}
\affil[4]{\orgdiv{Astrophysics}, \orgname{Department of Physics, University of Oxford}, \orgaddress{\street{Parks Rd}, \city{Oxford}, \postcode{OX1 3PU}, \country{UK}}}
\affil[5]{Steward Observatory, Department of Astronomy, University of Arizona, Tucson, AZ 85719, USA}
\affil[6]{\orgdiv{Institut Trottier de Recherche sur les Exoplanètes et Département de Physique}, \orgname{Université de Montréal}, \orgaddress{\street{Av. Thérèse-Lavoie-Roux}, \city{Montréal}, \postcode{H2V 0B3}, \country{Canada}}}
\affil[7]{Department of Astronomy \& Astrophysics, University of California, Santa Cruz, CA 95064, USA}

\affil[8]{Anton Pannekoek Institute for Astronomy, University of Amsterdam, Science Park 904, 1098 XH, Amsterdam, the Netherlands}

\affil[9]{Tsung-Dao Lee Institute \& School of Physics and Astronomy, Shanghai Jiao Tong University, Shanghai 201210, People’s Republic of China}

\affil[10]{Department of Astronomy \& Astrophysics, University of Chicago, Chicago, IL 60637, USA}
\affil[11]{Leibniz Institute for Astrophysics Potsdam, An der Sternwarte 16, 14482 Potsdam, Germany}

\affil[12]{Earth and Planets Laboratory, Carnegie Institution for Science, Washington, DC 20015, USA}
\affil[13]{\orgdiv{Atmospheric, Oceanic, and Planetary Physics}, \orgname{Department of Physics, University of Oxford}, \orgaddress{\street{Parks Rd}, \city{Oxford}, \postcode{OX1 3PU}, \country{UK}}}
\affil[14]{Department of Astronomy, University of Maryland, College Park, MD 20742, USA}
\affil[15]{DESY, Platanenallee 6, D-15738 Zeuthen, Germany}


\abstract{Hot Jupiters have temperature gradients of several hundreds of degrees between their permanent day and nightsides. In equilibrium, the primary carbon reservoir is expected to transition from CO on the dayside to CH4 on the nightside. Theory predicts that the atmospheric circulation, characterised by km/s winds, can advect chemical species from the dayside to the nightside faster than the time needed for the CO-to-CH4 chemical reaction to reach equilibrium. However direct evidence of this process has, so far, remained elusive, partly because it is often degenerate with other processes, such as vertical mixing or non-stellar elemental abundances. Here, we present observational evidence for such day-to-night transport of chemical species by observing both the dayside and the nightside of the hot Jupiter NGTS-10A b with the JWST/NIRSpec instrument. We constrain the presence of H2O and CO with similar abundances on both the dayside and nightside. Our observations are compatible with a solar-composition atmosphere at chemical equilibrium on the dayside, but indicative of disequilibrium chemistry for the nightside as it is significantly depleted in CH4 compared to equilibrium chemistry predictions. We further show that the lack of CH4 on the planet’s nightside cannot be attributed to non-solar elemental abundances or to vertical mixing mechanisms and must therefore be due to horizontal chemical quenching. Our study shows the fundamental role atmospheric transport plays in shaping the distribution of chemical species on exoplanet atmospheres.}
\newpage

%%================================%%
%% Sample for structured abstract %%
%%================================%%

\keywords{Exoplanets, atmospheres, JWST, Hot Jupiter}



\maketitle
\newpage
\newpage

\section*{Main}\label{sec:main}
Among the thousands of known exoplanets, hot Jupiters provide new opportunities to test our knowledge of atmospheric physics. Their tidally-locked rotation and the strong irradiation they receive from their parent star lead to a strong contrast between their hot dayside and their cooler nightside. In turn, this fuels kilometer-per-second winds that transport energy, clouds, and chemical species, driving the atmosphere out of local equilibrium~\citep{Showman2020}. 

Measurements with previous space telescopes, such as Spitzer and Hubble, have provided direct evidence of heat transport by the winds by measuring the nightside temperatures and the displacement of the hottest point of the atmosphere away from the substellar point for a sample of these objects~\citep[][]{knutson2007,Parmentier2018a,Bell2021,Sikora2025}. More recently, the direct measurement of winds has become possible from ground-based high-spectral resolution instruments of the hottest of these objects~\citep{Gandhi2022,simonninTimeResolvedAbsorption2024}. Horizontal transport of chemical species has been postulated as a fundamental process that determines the chemical abundances in hot Jupiter atmospheres but has not yet been firmly observed.

We observed NGTS-10A b within JWST GO program 2158 (PI: Parmentier). NGTS-10A b was discovered by the NGTS consortium~\citep{McCormac2020} and is one of the shortest period hot Jupiters, completing an orbit in 18.4h around a relatively cool \mbox{K5V} and dim \mbox{$K_{\rm mag}=11.8$} star. It orbits extremely close to its host star (with a semi-major axis $1.46\pm 0.18$ Roche radii), meaning that it is expected to spiral down and be engulfed by its host star in the next tens of Myr~\citep{McCormac2020}. We specifically chose this short-period planet around a dim star so that it could be monitored throughout its whole orbit using the NIRSpec/PRISM instrument mode without saturating the detector. The planet sits in a temperature range that is a perfect test case to study the transport of chemical species through carbon chemistry. As we will show, the main carbon carrier is expected to be carbon monoxide on the dayside of the planet, whereas the nightside of the planet is expected to be cold enough for methane to become dominant.

\section{Results}
\subsection{Thermal phase curve}
\begin{figure} 
    \centering
    \includegraphics[width=\linewidth]
    {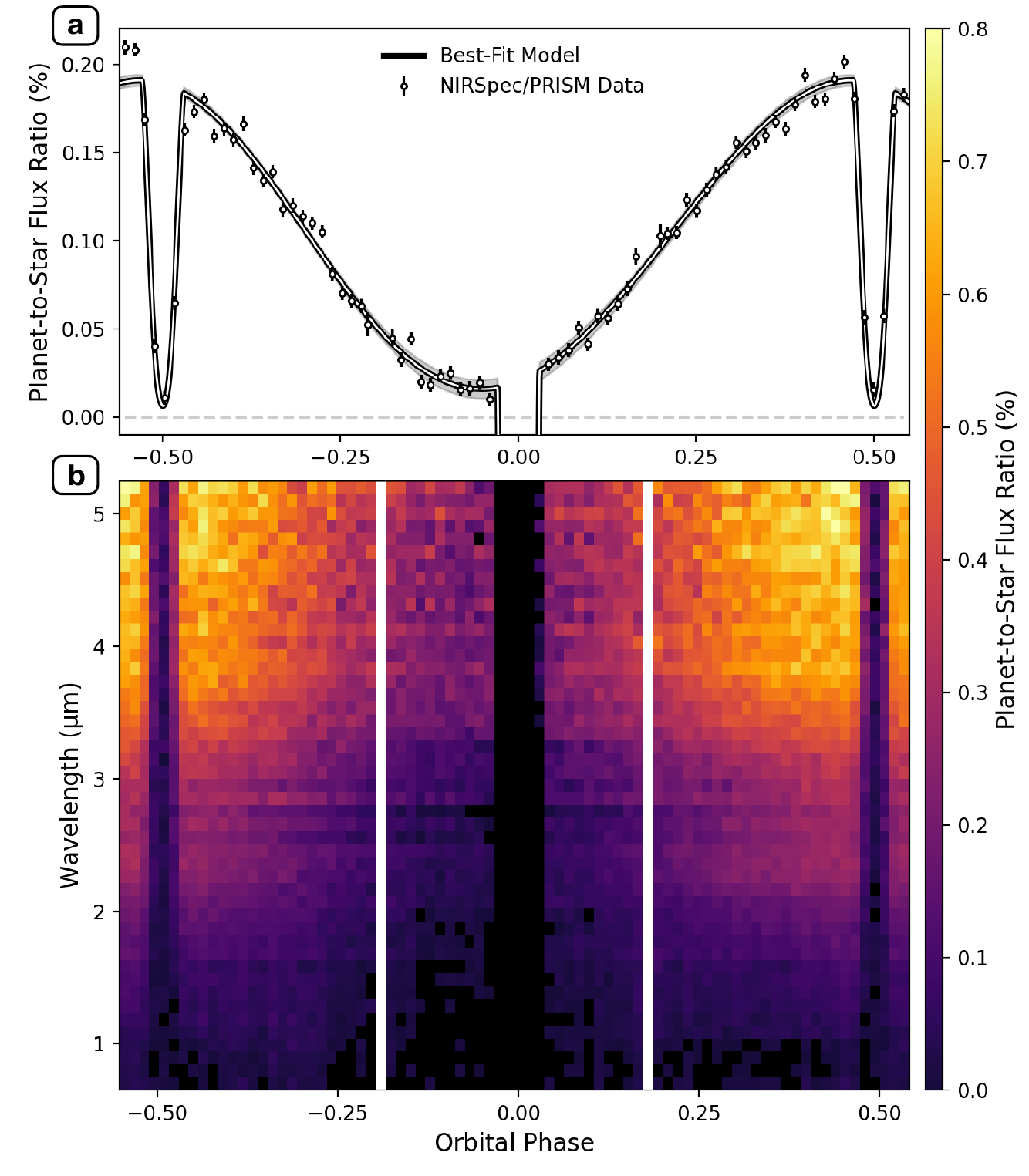}
    \caption{{\bf JWST/NIRSPEC/PRISM orbital phase curve observation of NGTS-10A b}{\bf a}, Band-integrated (white-light) phase curve of NGTS-10A b. Values above the  dashed line indicate thermal emission from the planet. Owing to the grazing  nature of the transit, the eclipse does not reach zero flux at orbital phases of $\pm0.5$. The planet transits the star at an orbital phase of zero. The NIRSpec/PRISM data (mean and standard deviation as white circles) are binned in time for  illustration purposes. The best-fitting phase model (black/white curve) includes a $1\sigma$ uncertainty region (in grey). {\bf b}, Spectroscopic phase curves of NGTS-10A b. The data were binned into the 47 spectroscopic channels shown here before  each phase curve was fitted (not shown, but like a). The white stripes depict the  masked regions at the start of each JWST exposure where there are noticeable systematic ramps. We detected thermal emission from the planetary nightside at wavelengths longer than $\approx 1.5\mu m$.
    }
    \label{fig:data}
\end{figure}

We observed the full orbital phase curve of the planet from 0.6 to 5.3 $\mu$m, beginning shortly before secondary eclipse and ending shortly after the subsequent eclipse. The observations started on 11 March 2023 and lasted 20.4 hours. We used the \texttt{Eureka!} pipeline~\citep{Bell2022} to reduce the data and generate 47 spectroscopic phase curves (Figure~\ref{fig:data}). We found an unexpected flux offset between each of the three exposures (see~\ref{fig:pc-ch38}). The offset is wavelength dependent (see ~\ref{fig:fluxOffsets}) and we followed by a decaying ramp we decided to mask (see \ref{fig:wlc}). We confirm that the transit and eclipse are grazing, with an impact parameter of $0.844 {\pm} 0.005$ stellar radii (see sect ~\ref{sec:fitting}). This implies that $96.4 {\pm} 0.9\%$ of the planetary disk is occulted at the secondary eclipse mid-point. We find that the NGTS-10 system is composed of two stars, NGTS-10A and NGTS-10B, separated by $\sim0.33\arcsec$ (see~\ref{fig:TA}) (see~\ref{fig:TA}). NGTS-10B is not the same star as that reported by\citep{McCormac2020} (G-6880), which is located $\sim1.2\arcsec$ away. We verify that the planet indeed orbits NGTS-10A by masking the light of NGTS-10B. The stars are separated by two pixels in the direction perpendicular to the dispersion direction of the instrument; however, we could not extract non-overlapping spectra for NGTS-10A and NGTS-10B (see \ref{fig:3304}). Instead, we used a phoenix model grid to jointly fit both spectra (see \ref{fig:stellar-fit}). We find that NGTS-10A is a K dwarf of effective temperature $4567\pm36$\,K with a radius of $0.70\pm0.03\,R_{\rm sun}$ while NGTS-10B is a $3670\pm140$\,K, $0.40\pm0.03\,R_{\rm sun}$ M dwarf. We measure similar distances for both stars. This indicates that they are likely part of the same system, with a projected separation of 107 AU, situated $346\pm 18 $\,pc from Earth {(see methods and~\ref{table:1})}.

This, together with our newly derived $a/R_\mathrm{s}=4.587\pm{0.011}$ correspond to an equilibrium temperature for full redistribution and zero albedo of $1508\pm12$\,K for the planet NGTS-10A b (see ~\ref{tab:white_parameters}).

\subsection{Heat transport}

\begin{figure*}
    \centering
    \includegraphics[width=\linewidth]{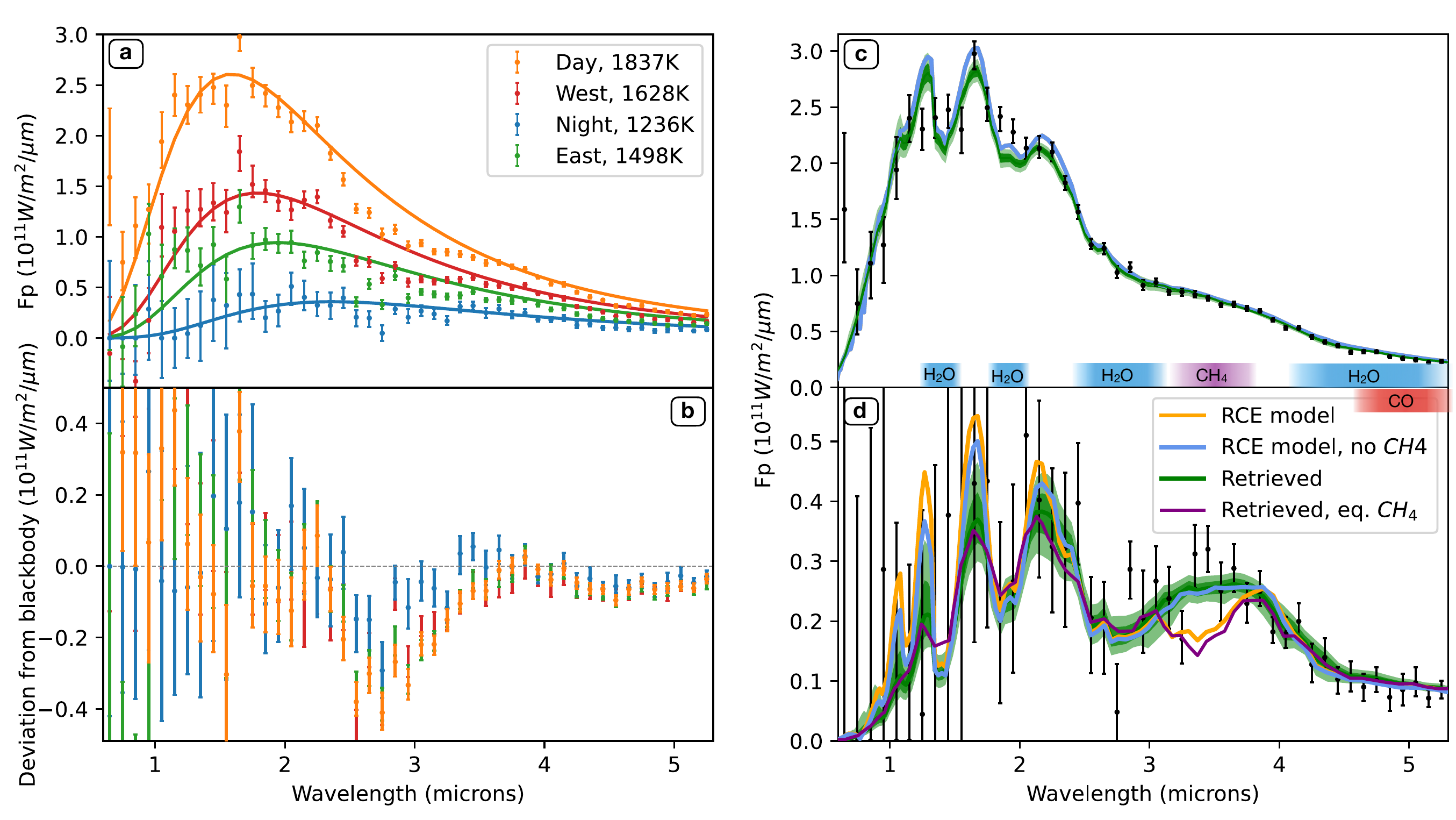}

    \caption{{\bf NGTS-10A b spectra.} {\bf a)} NGTS 10A b planetary spectra at four phases (mean and standard deviation) compared to blackbody with the same total outgoing flux (see text). {\bf b)} deviation of the spectra from the equivalent blackbodies. A broad absorption band from water at 2.8 $\mu$m, along one from CO and CO$_2$ at 4.5 $\mu$m, are clearly visible. The position of these spectral bands together with their strength and shape is unchanged for all phases, implying that the chemical composition is not changing significantly over the whole planet. Particularly, no \ce{CH4} bands appear in the 3-4$\mu m$ range at all phases.\\
    {\bf c)}  Dayside and {\bf d)} nightside spectra of NGTS-10A b compared to our best fit retrieved spectrum (1- and 2-$\sigma$ ranges in green) and our radiative/convective models in chemical equilibrium (orange) and with a depletion of \ce{CH4} (blue). We also show the best fit model from our free retrievals with \ce{CH4} added in equilibrium abundances (purple). Whereas the chemical equilibrium model is a good match on the dayside, \ce{CH4} appears to be depleted on the nightside compared to equilibrium values. For the dayside, models with and without \ce{CH4} overlap perfectly.}
    \label{fig::Deviation}
\end{figure*}

By computing the mean and standard deviation of the phase curve model component for each wavelength\citep[e.g.,][]{Stevenson2014a}, we extracted the planetary emission spectrum at four different orbital phases. As usual with exoplanets that are not spatially resolved, we rely on differential measurements in time to extract the planetary spectra. The data leads to a relative measurement of the planetary and stellar flux, which can be converted into a direct measurement of planetary flux by using the best-fit stellar spectra observed during the two eclipses of the planet taking into account the grazing geometry. As shown in Figure ~\ref{fig::Deviation}, our 0.6-5.3 $\mu$m observation covers most of the spectral energy distribution (SED) of the planet at all phases. By integrating this SED and matching the total flux to an equivalent blackbody we obtain a dayside effective temperature of $1837\pm10$\,K while the nightside is 600~K cooler, at $1236\pm27$\,K. These temperatures correspond to a heat redistribution efficiency of $\epsilon= 0.34\pm0.02$~\citep{Cowan2011} (where 1 is full redistribution and 0 is no heat transport), implying a rather poor heat redistribution compared to planets with the same equilibrium temperature\citep{May2022} and pointing towards the presence of nightside clouds or drag mechanism reducing the day-to-night heat transport\citep{Parmentier2021,Roth2024}. The same set of values, using the formalism of \citep{Cowan2011} leads to a Bond Albedo of $A_{\rm B}= -0.14^{+0.08}_{-0.02}$. This value is negative, but compatible with 0 as the $2-\sigma$ level, pointing towards a small Bond albedo for the planet, and thus the lack of a reflective dayside cloud cover. The eastern hemisphere temperature is $1628^{+10}_{-4}$\,K, significantly hotter than the western one, at $1481^{+12}_{-5}$\,K. The east/west asymmetry is typical of hot Jupiter atmospheres and indicative of an atmospheric circulation dominated by strong, eastward, super-rotating winds \citep{Showman2020}.

\subsection{Chemical composition}

\begin{figure}
    \centering
    \includegraphics[width=\linewidth]{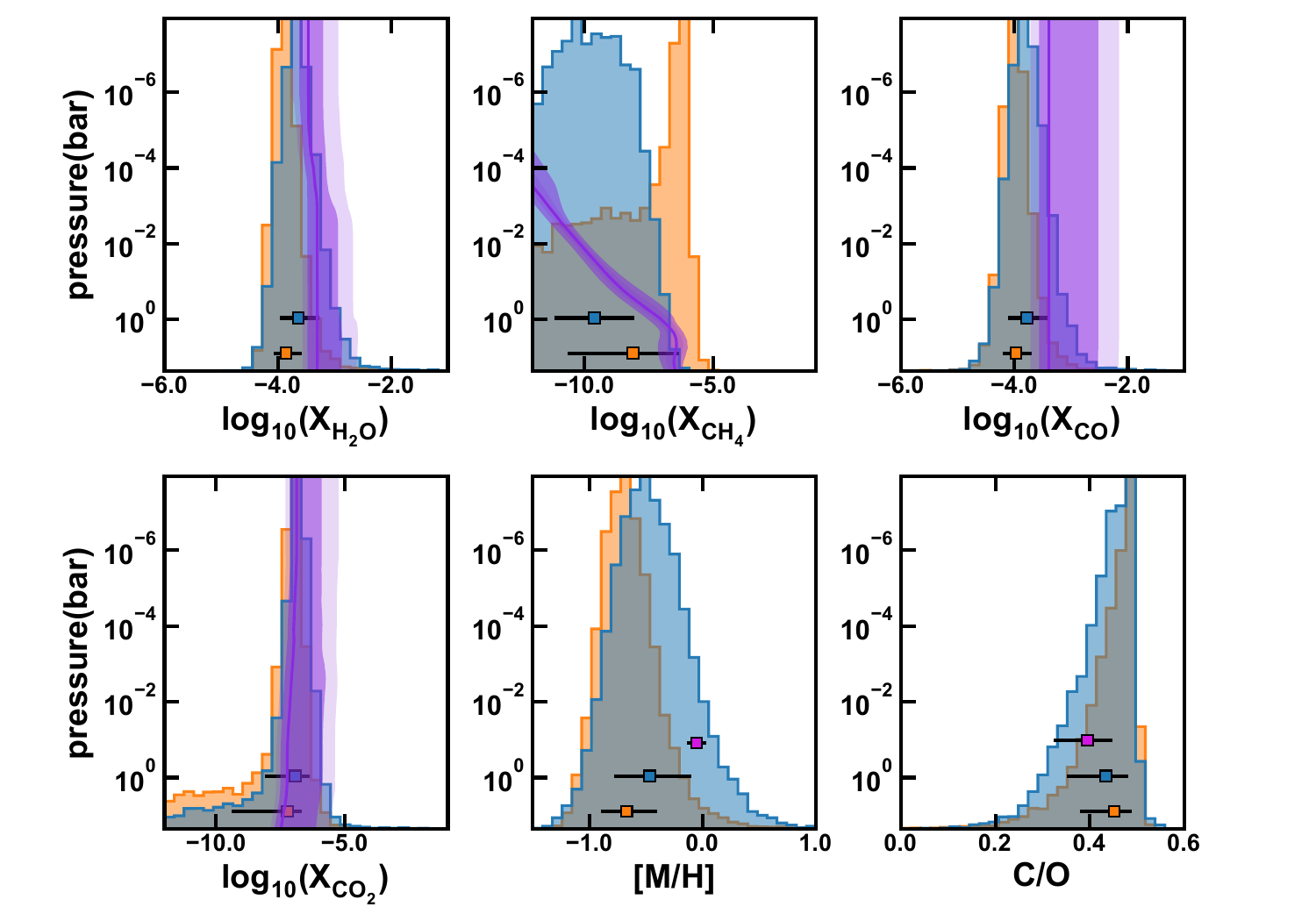}
    \caption{{\bf Chemical composition of NGTS-10A b.}Chemical abundances of the major species on the day (orange histograms) and night (blue histograms) sides of NGTS-10A b obtained from free retrievals. The points show the mean and the $1-\sigma$ spread of the histograms and are placed at an arbitrary vertical position. Our retrieved abundances of \ce{H2O} and \ce{CO} are similar in both sides of the planet. The grid retrieval under chemical equilibrium applied to the dayside atmosphere {shows chemical abundances consistent within $2-\sigma$ with the free retrieval approach, both for the chemical abundances (solid lines with 1 and $2-\sigma$ confidence intervals as shaded area) and the bulk abundance (purple squares with $1-\sigma$ confidence interval for metallicity and C/O)}. {The nightside \ce{CH4} abundance $3-\sigma$ upper limit of $10^{-6.5}$ is much smaller than the expected nightside abundance at equilibrium ($10^{-5.2}$})
    }
    \label{fig::retrieval}
\end{figure}

Panel b of Fig~\ref{fig::Deviation} shows the difference between the observations and the blackbody at the calculated effective temperature for our four phases. Our observations show clear departures from blackbody at all phases, particularly in the broad water band at $2.5-3.5\mu$m and in the $4-5\mu$m region where we expect absorption from CO and \ce{CO2}. Importantly, the position of these spectral bands together with their strength and shape is unchanged for all phases, implying that the chemical composition is not changing significantly over the whole planet. Particularly, no \ce{CH4} bands appear in the 3-4{\microns} range at any phase. The presence of higher uncertainties (\ref{fig:scatter_mult}) and correlated noise (see \ref{fig:CorrNoise}) prevents us from detecting the water bands at wavelengths shorter than $>1.7$ {\microns} In order to verify that the composition is constant with phase, we use the Aurora~\citep{Welbanks2021, Bell2023} free retrieval framework to estimate the chemical volume mixing ratios (which we later on call abundances) and thermal structure of the day and nightside hemispheres of the planet. As shown in panels c and d of Figure~\ref{fig::retrieval}, we constrain \ce{H2O} and CO on both hemispheres, and find only upper limits on the other molecules. Our dayside and nightside water abundances are consistent within 1$\sigma$ ($\log_{10}($H$_2$O$_{\rm day})=-3.31^{+1.14}_{-0.56}$ and $\log_{10}($H$_2$O$_{\rm night})=-3.63^{+0.60}_{-0.32}$ respectively). Similarly, the CO abundances are consistent between the two hemispheres ($\log_{10}($CO$_{\rm day})=-3.80^{+0.92}_{-0.71}$ and $\log_{10}($CO$_{\rm night})=-3.78^{+0.57}_{-0.34}$ respectively). Furthermore, the free retrieval allows us to put upper limits on \ce{CH4}, \ce{NH3}, and \ce{CO2} abundances on both the dayside and the nightside. Particularly, the nightside retrieval leads to a 3-$\sigma$ upper limit of $10^{-6.5}$ for the mixing ratio of \ce{CH4}.

To confirm the physical plausibility of our retrieved abundances, we use the ScCHIMERA framework \citep{Line2013a} to make a grid of atmospheric models in radiative/convective/chemical equilibrium for the dayside atmosphere. We vary irradiation \citep[to marginalize over the possible values of Bond albedo and recirculation efficiency,][]{Arcangeli2018}, C/O ratio, and metallicity. Through a nested sampling algorithm, we find that the best fit model has close to a solar metallicity ($[\text{M/H}]=-0.05^{+0.04}_{-0.07}$]) and a slightly subsolar C/O ratio ($\text{C/O}=0.39^{+0.05}_{-0.07}$). As shown by the colored lines of Fig.\ref{fig::retrieval}, the mean chemical abundances from the radiative/convective/chemical equilibrium dayside retrieval are within 2$\sigma$ of the chemical abundances obtained from the free retrieval framework, meaning that the dayside atmosphere is consistent thermochemical equilibrium at the pressure levels probed by the JWST observations. However, the fact that the free chemistry approach retrieves less water and CO than the chemical equilibrium leads to a lower metallicity estimate from the free retrieval approach but a similar C/O ratio ($[M/H]=-0.46_{-0.31}^{+0.36}$ and $C/O=0.45_{-0.07}^{+0.037}$) on the dayside, and $[M/H]=-0.67_{-0.22}^{+0.26}$ and $C/O=0.43_{-0.08}^{+0.05}$ on the nightside).

When applied to the nightside, the chemical equilibrium radiative/convective retrieval converges towards a higher metallicity solution ($[M/H]= 0.83\pm0.36$, $C/O=0.42\pm0.15$). This higher metallicity solution is in contradiction with the free retrieval approach which finds similar metallicity and C/O ratio on the dayside and the nightside. It is further in contradiction with our expectation that the bulk elemental abundances should not change drastically throughout the atmosphere. The higher metallicity retrieved from the nightside is likely a bias due to the chemical equilibrium assumption, which is assumed in the radiative/convective modelling but not in the free retrieval. Indeed, over this temperature range, increased metallicity disfavors \ce{CH4}~\citep{Lodders2002a}, causing the radiative/convective/chemical equilibrium framework to increase the metallicity to reduce \ce{CH4}absorption.

\section{Discussion}

\begin{figure}
    \centering
    \includegraphics[width=\linewidth]{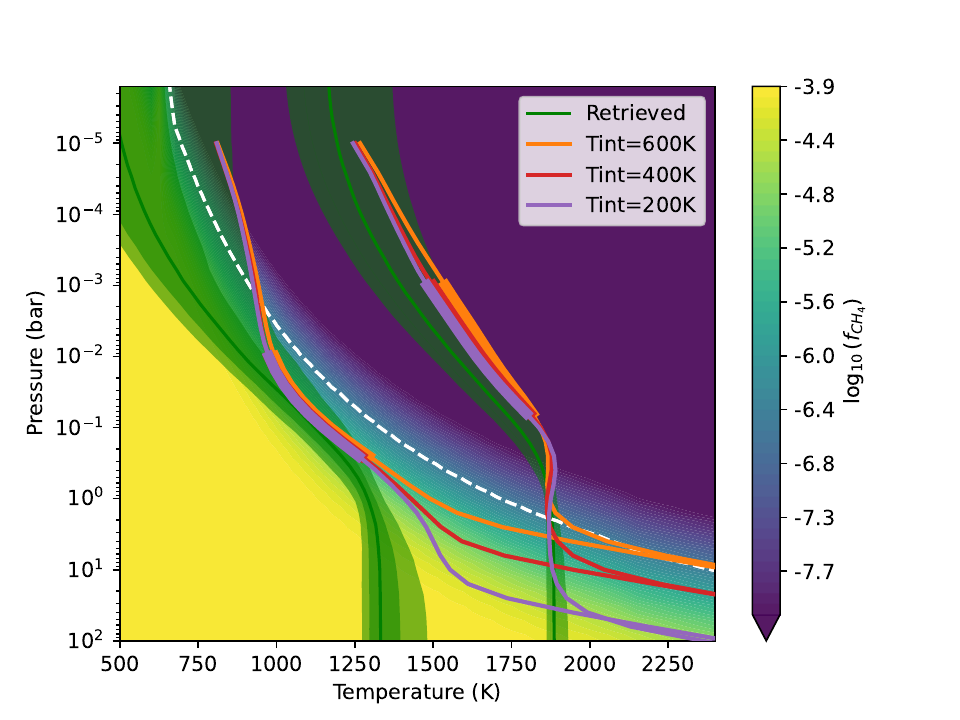}
    \caption{{\bf Thermochemical structure of NGTS-10A b.} Chemical abundance of \ce{CH4} as a function of pressure and temperature { assuming chemical equilibrium for our dayside retrieved elemental abundances ([M/H]=-0.05  and C/0=0.39). The retrieved thermal structure of the planet dayside is shown in green, together with different radiative/convective models obtained by adjusting the irradiation flux to match the total outgoing flux measured. The internal temperature corresponds to the one of the dayside. The nightside intrinsic temperature is higher by 200K to match the same deep thermal profile as the dayside. The photospheric regions probed by the observations is indicated by the thicker lines. The hotter sets of profiles correspond to the dayside whereas the cooler set of profiles correspond to the nightside.}The white dashed line shows the $3-\sigma$ upper limit of \ce{CH4} derived from the nightside observations. We see that chemical equilibrium predicts large enough abundance of \ce{CH4} on the nightside to be detectable but not on the dayside. In order for vertical quenching to deplete the nightside, we would need a $T_{\rm int}>600$\,K, which is ruled out by the measured mass and radius. Horizontal transport from day-to-night can, however, deplete the nightside of \ce{CH4}. 
    }
    \label{fig::CH4}
\end{figure}

We can use the metallicity and C/O ratio derived from the dayside spectrum to estimate the expected \ce{CH4} abundance on the nightside if it were in chemical equilibrium. As shown in Fig.~\ref{fig::CH4}, the predicted nightside abundance of \ce{CH4} at photospheric pressures is $\approx10^{-5.2}$. This is $10\times$  higher than the upper limit allowed by the data. To verify this, we ran a radiative convective models at chemical equilibrium and one where we artificially set the \ce{CH4} abundance to zero. As shown by the orange and and blue lines of figure~\ref{fig::Deviation}, an equilibrium abundance of CH4 on the nightside is ruled out by the shape of the spectrum between 3 and 4\,$\mu$m. We further confirmed this by artificially adding \ce{CH4} to the retrieved spectrum (purple line) to mimick an equilibrium chemistry. We therefore conclude that whereas the dayside atmospheric molecular abundances are consistent with thermochemical equilibrium, the nightside atmosphere is not in chemical equilibrium. 

The nightside composition of NGTS-10A b can be driven out of chemical equilibrium by two main processes: vertical mixing and horizontal mixing. Both processes result from the same physics: a competition between chemical timescales and transport timescales. Particularly, the chemical timescale decreases strongly with increasing temperature. Thus at a given pressure level, the hotter dayside of the atmosphere can more quickly reach chemical equilibrium, whereas the cooler nightside can have a chemical composition imposed by the hotter dayside if the atmospheric circulation mixes the chemical composition faster than it would take for chemistry to evolve the composition in the cooler parts towards equilibrium.
Disequilibrium carbon chemistry driven by vertical mixing has been observed in Jupiter ~\citep{Beer1975,Fegley1988}, brown dwarfs~\citep{Mukherjee2022} and directly imaged planets~\citep{Molliere2020}. More recently, similar detection of vertical quenching was obtained on exoplanets with JWST~\citep{Welbanks2024}. It relies on the idea that the chemical timescale becomes shorter at depth, due to both increased temperature and increased pressure. So, if vertical movements in the atmosphere are fast enough, the chemical composition at the photosphere can be determined by the conditions deeper than the photosphere~\citep{Visscher2011}. Previous detailed chemical models of WASP-43b~\citep{Venot2020}, a planet very similar to NGTS-10Ab, show that the quench level should be around 1 bar. It further shows that, for a cold interior model, vertical mixing is unable to reduce the \ce{CH4} abundance to values lower than $10^{-6}$. For vertical mixing to reduce the \ce{CH4} abundance in NGTS-10A b below our upper limit of $10^{-6.5}$, it would be necessary for the nightside temperature profile of the planet to become hot enough to reach the zone in the right side of the dashed line of Fig. \ref{fig::CH4}, where CO dominates over \ce{CH4}. As can be seen in Figure~\ref{fig::CH4}, this would happen only if the internal flux of the planet corresponds to a temperature { significantly} greater than $600$\,K. {We used the well established interior modeling framework of~\cite{Fortney2007} to estimate the radius of NGTS-10b for different values of internal temperature and core mass. As shown in \ref{fig::interior}, the measured radius of NGTS-10b is only compatible with an intrinsic temperature below $500$\,K for a reasonable range of core masses $(0-60\,M_{\rm Earth})$.} As a consequence, vertical quenching of the nightside abundances is not a plausible mechanism to explain the lack of methane.

A second mechanism to drive the nightside atmosphere out of chemical equilibrium is horizontal transport. If the atmospheric winds transport the chemical species from the hot, \ce{CH4}-poor dayside to the nightside fast enough, then the nightside should also be devoid of \ce{CH4}. As shown in Figure~\ref{fig::CH4}, the equilibrium abundance of \ce{CH4} on the dayside is $10^{-10}$, which would be completely undetectable both on our dayside and nightside spectra. Nightside disequilibrium chemistry due to horizontal transport has been postulated since 2006~\citep{Cooper2006} and is possible only if the horizontal wind speeds at the photosphere are fast enough compared to the chemical timescale. At 1~bar, the CO/\ce{CH4} interconversion timescale is $\approx 10^{7}$\,s at $1500$~K and $\approx 10^{10}$~s at $1000$~K~\citep{Visscher2012}. { Given the circumference of NGTS-10b, gas will be transported horizontally from one hemisphere to the other within such timescales so long as wind speeds are larger than a few tens of m/s}. Wind speeds predicted by global circulation models for hot Jupiters with similar temperatures are  generally more than 10 times faster than this, on the order of km/s~\citep{Kataria2015,Roth2024}.

{ Recently the non-detection of \ce{CH4} on the nightside of the hot Jupiter WASP-43b observed with JWST/MIRI~\citep{Bell2024} has been interpreted has a tentative evidence of such a transport. However, the study cannot provide a firm conclusion regarding horizontal quenching of \ce{CH4}. First, because they could not capture the full spectral energy distribution of the planet, and the MIRI wavelength range lacks the presence of strong CO bands, the bulk atmospheric composition was not determined well enough to narrow the range of possible \ce{CH4} abundance expected at equilibrium (1–100 ppm). Second, the low SNR of the MIRI spectra led to a lower-limit on the  \ce{CH4} abundance of 10 ppm that is too large to conclude robustly about the presence of disequilibrium chemistry.

By capturing the full spectral energy distribution of a NGTS-10A b at all phases, our observations are able to constrain both the temperature and the chemistry at multiple orbital phases, { raising the degeneracies present in the WASP-43b observations}. This allows us to show observationally that the carbon chemistry of hot Jupiters can be homogenized and driven out of chemical equilibrium by the atmospheric circulation. { Our result is in agreement with the tentative conclusion of horizontal quenching for the hot Jupiter WASP-43b~\citep{Bell2024}, but is in contrast with the recent detection of \ce{CH4} in the nightside of the ultra-hot Jupiter WASP-121b~\citep{Evans-Soma2025}. We propose two reasons that can explain why horizontal quenching determines the \ce{CH4} abundance on NGTS-10b but not on WASP-121b. First, the nightside of WASP-121b is hotter by $\approx 250$\,K compared to the one of NGTS-10b. This would reduce the chemical timescales by two orders of magnitudes \citep{Visscher2012}, favouring the equilibrium value. We note, that, despite the hotter nightside temperatures seen on WASP-121b, the higher C/O ratio of the planet atmospheres (0.92 vs. 0.39 in NGTS-10b) favours the presence of \ce{CH4} on the planet nightside. An additional possibility is that wind speeds on WASP-121b could be smaller than in NGTS-10b due to the presence of magnetic drag acting on ultra-hot Jupiter flows \citep{Beltz2022}. This, combined with the $50\%$ higher radius of WASP-121b could increase the day-to-night transport timescale, favouring the presence of equilibrium \ce{CH4} abundances on the nightside of WASP-121b. }

Our results show that transport-driven chemistry can play an important part in setting the chemical composition of exoplanet atmospheres. NGTS-10A b will serve as a benchmark  for future models coupling atmospheric circulation and chemistry that can then be applied to a wide range of exoplanets.

\bmhead{Acknowledgements}
{ We are grateful for the long and careful reports from the three referees, that have strengthen the conclusions of our work. 
This work was partially funded
by the French National Research Agency (ANR) project EXOWINDS (ANR-
23-CE31-0001-01). This work is based on observations made with the NASA/ESA/CSA James Webb Space Telescope. The data were obtained from the Mikulski Archive for Space Telescopes at the Space Telescope Science Institute, which is operated by the Association of Universities for Research in Astronomy, Inc., under NASA contract NAS 5-03127 for JWST. These observations are associated with program $\#2158$. J.T. was supported by the Glasstone Benefaction,
University of Oxford (Violette and Samuel Glasstone Research Fellowships in Science 2024).
J.M.D acknowledges the research program VIDI New Frontiers in Exoplanetary Climatology with project number 614.001.601, which is (partly)
financed by the Dutch Research Council (NWO)
}

\section*{Author contribution}
VP led the original proposal, performed the 1D radiative/convective modeling and wrote the manuscript. 
KBS reduced the observations and contributed to the manuscript writing. 
LW performed the free atmospheric retrievals used in the article.
JT performed initial atmospheric retrieval explorations. 
ES contributed to the proposal and performed an initial data reduction.
L-PC contributed to the proposal text and the effective temperature calculations.
YT performed the internal modeling of NGTS-10b.
ML performed the grid retrievals.
HS performed a preliminary fit of the stellar spectra. 
XT  provided insights on the transport properties of NGTS-10b through GCM modeling. 
MH performed an initial fit of the lightcurves.
VP, JB, J-MD, JF, PG, TK, ML, ES, KS, JT, EK, MM were all part of the original JWST proposal. 
All authors commented on the manuscript draft. 

\noindent The authors declare no competing interests.

\noindent Correspondence and requests for materials should be addressed to Vivien Parmentier. 
%\bibliography{Biblio_NGTS10,references}% common bib file
\bibliography{references}% common bib file

%%%%%%%%%%%%%%%%%%%%%%%%%%%%%%%%%%%%%%%%%%%%%%%%%%%%%%%%%%%%%%%%%%%%%%%

\clearpage
\section*{Methods} \label{sec:methods}
\renewcommand{\figurename}{Extended Data Fig.}
\renewcommand{\tablename}{Extended Data Table}
\renewcommand{\thefigure}{Extended Data Fig.~\arabic{figure}}
\renewcommand{\theHfigure}{Extended Data Fig.~\arabic{figure}}
\renewcommand{\thetable}{Extended Data Table \arabic{table}}
\renewcommand{\theHtable}{Extended Data Table \arabic{table}}
\setcounter{figure}{0}
\setcounter{table}{0}

\subsection*{Data Reduction} \label{sec:reduction}
% I guess we need both Everett and Kevin's reduction here ?

We used the NIRSpec/PRISM instrument mode to observe the NGTS-10 systems for 20 contiguous hours.  Due to its faintness, we were able to perform target acquisition on NGTS-10A using the F110W filter.  Upon inspection of the data, we found a nearby star that we label NGTS-10B (see \ref{fig:TA}).  We provide addition details in the section on Stellar Characterisation.

The science observation used six groups per integration, 15,513 integrations per exposure, and three exposures.  More than one exposure was required due to limitations on the number of frames per exposure (maximum of 196,608 frames).  We could have used two exposures; however, we had concerns about starting a new exposure during transit.  Our concerns were validated by the detection of wavelength-dependent flux offsets between exposures{ (see \ref{fig:pc-ch38} and \ref{fig:fluxOffsets}) }and small ramps at the start of each exposure (see \ref{fig:wlc}).  Each exposure lasted 6.8 hours.

\begin{figure}
    \centering
    \includegraphics[width=0.8\linewidth]{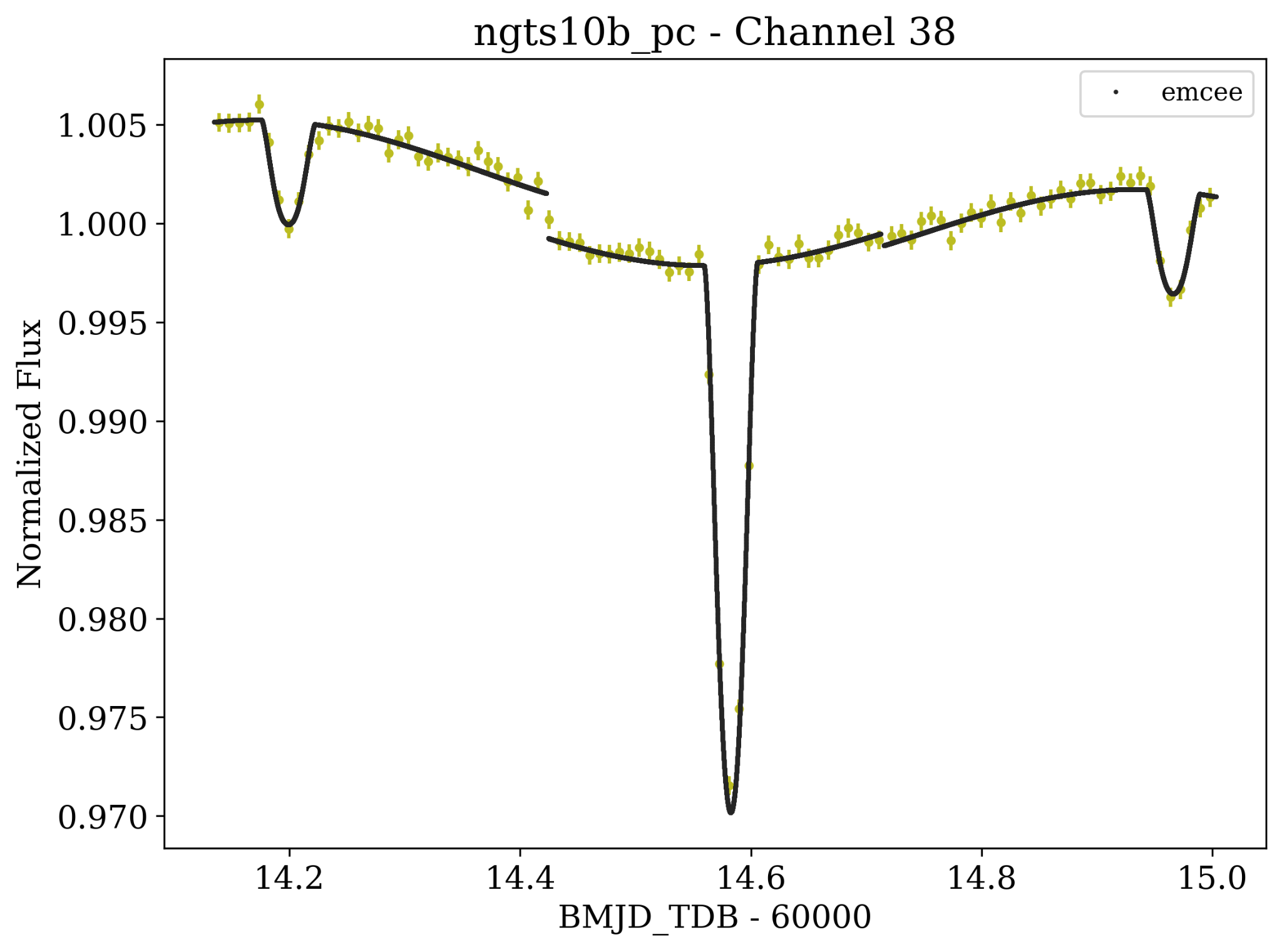}
    \caption{{\bf Observed lightcurve.} Spectroscopic phase curve of NGTS-10A b at 4.45 {\microns} showing mean and standard deviation of the flux.  The breaks between exposures at 14.4235 and 14.7140 days are made evident by the significant flux offsets.  The magnitude and direction of the flux offset varies between spectroscopic channels (see \ref{fig:fluxOffsets})}    
    \label{fig:pc-ch38}
\end{figure}

\begin{figure}
    \centering
    \includegraphics[width=0.8\linewidth]{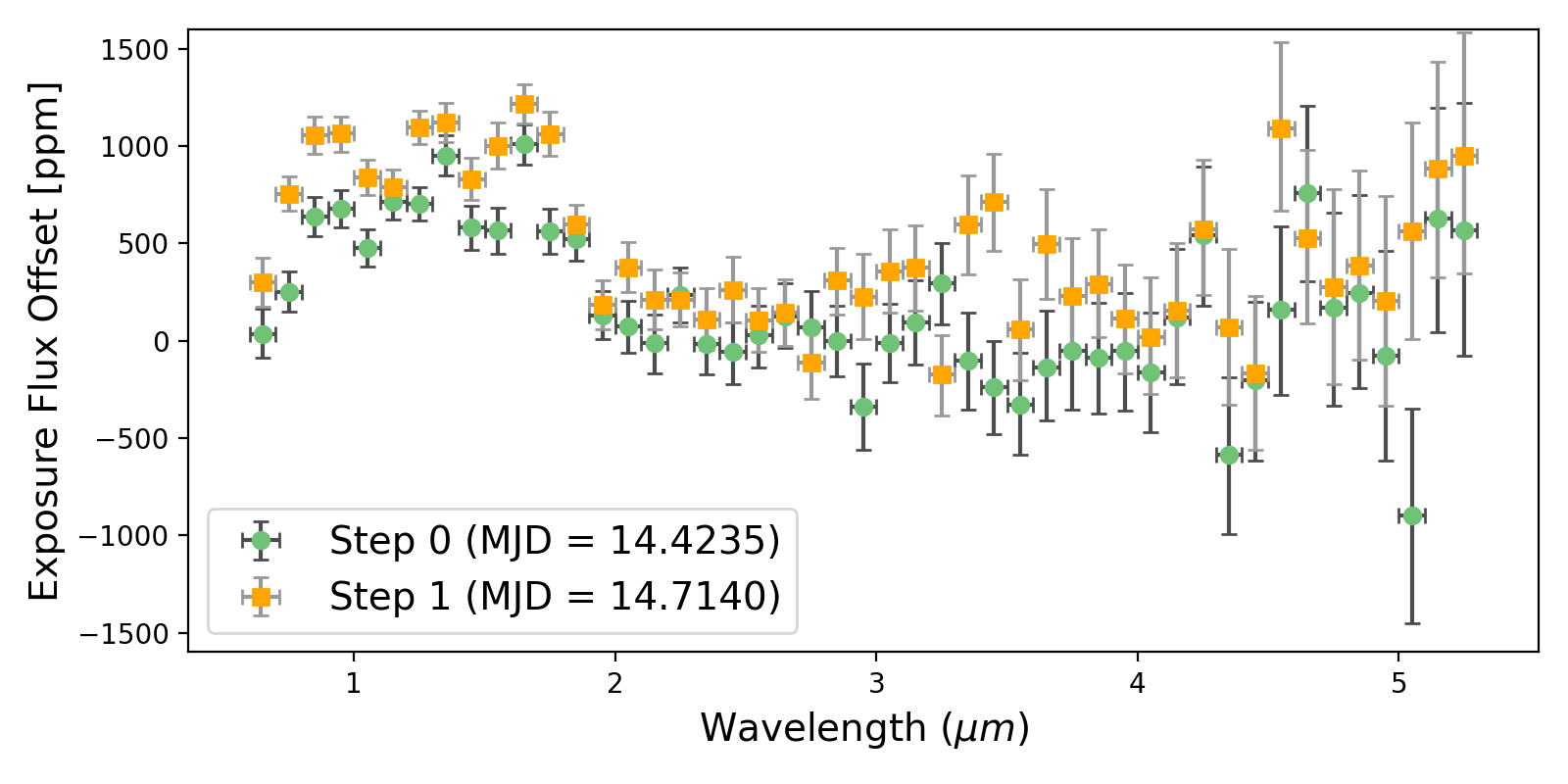}
    \caption{{\bf Flux offsets.} Measured flux offsets between our three exposures.  The green and orange points depict the measured flux offsets at MJD = 14.4235 and 14.7140, respectively with the $1-\sigma$ standard deviation as errorbars.}
    \label{fig:fluxOffsets}
\end{figure}

\begin{figure}
    \centering
    \includegraphics[width=0.8\linewidth]{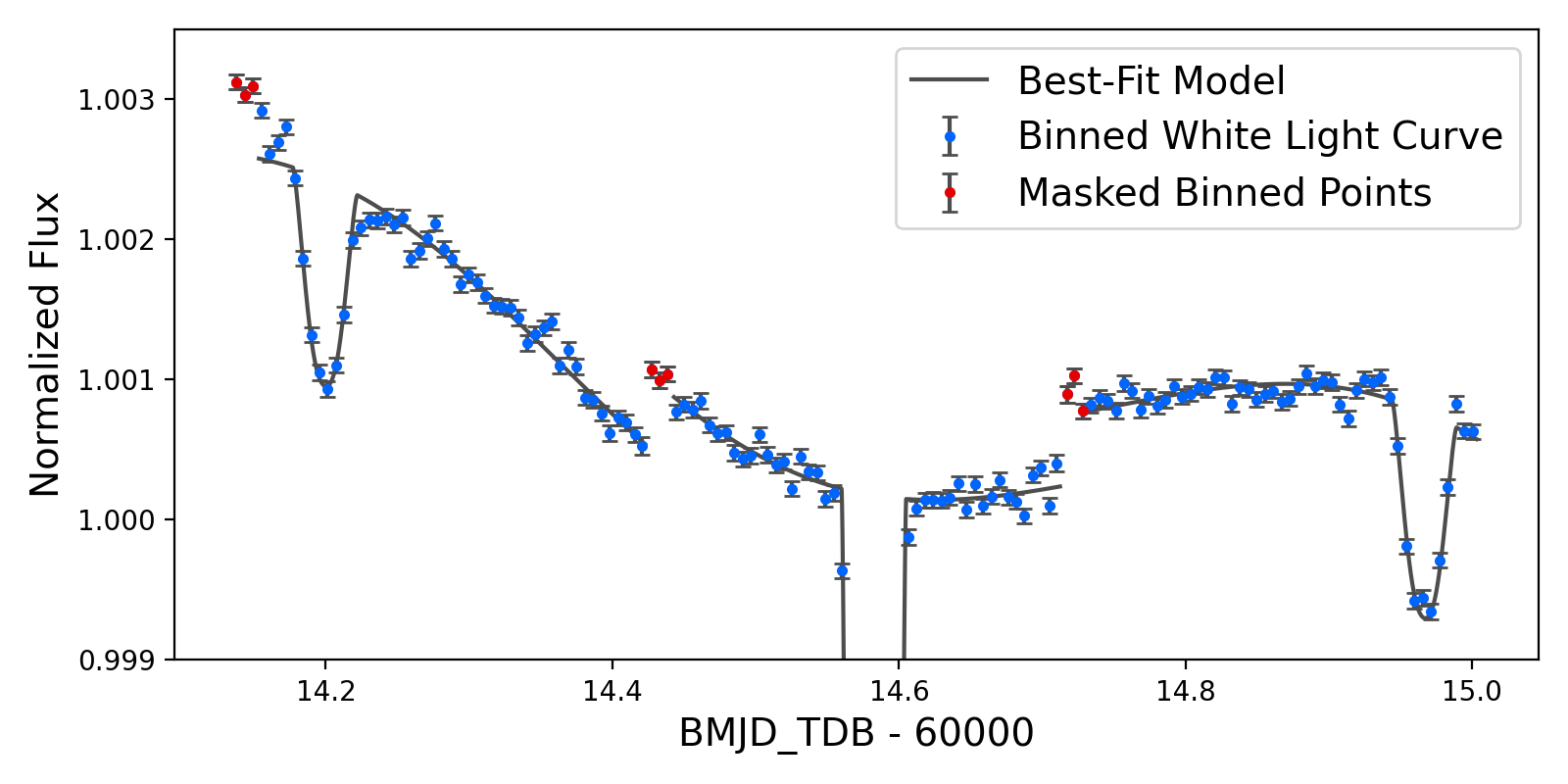}
    \caption{{\bf White lightcurve.}Binned white light curve (blue points with $1-\sigma$ errors) with best fit model (grey line).  The red binned points represent the 1,000 masked integrations at the start of each exposure that exhibit a decaying ramp systematic.}
    \label{fig:wlc}
\end{figure}

Starting from the uncalibrated FITS files, we used standard data reduction techniques via the \texttt{Eureka!} pipeline \citep{Bell2022} to generate white light and spectroscopic phase curves.  In Stage 1, we skipped the jump detection step and performed group-level background subtraction.  In Stage 2, we skipped the {\em flat\_field} and {\em photom} steps for the nominal reduction, but included them when generating the calibrated stellar spectrum.  In Stage 3, we used an aperture full-width of 3 pixels for the optimal spectral extraction step to minimize any contamination from NGTS-10B's spectrum (see \ref{fig:3304}).

\subsection*{Stellar Characterisation} \label{sec:fitting}
\begin{figure}
    \centering
    \includegraphics[width=0.7\linewidth]{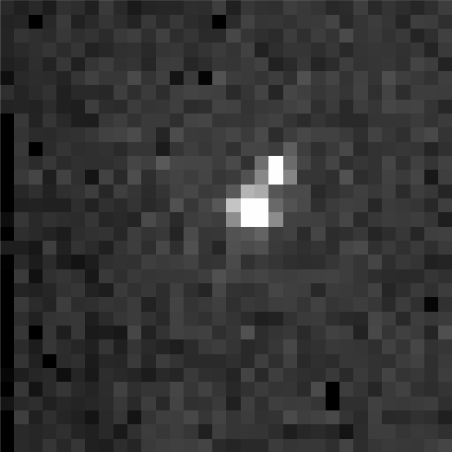}
    \caption{{\bf NGTS-10 system.} Target acquisition image using NIRSpec.  The brighter source near the center of the image is NGTS-10A; the fainter source up and to the right is NGTS-10B. The two stars are separated by $\sim0.33\arcsec$.}
    \label{fig:TA}
\end{figure}

\begin{figure}
    \centering
    \includegraphics[width=1\linewidth]{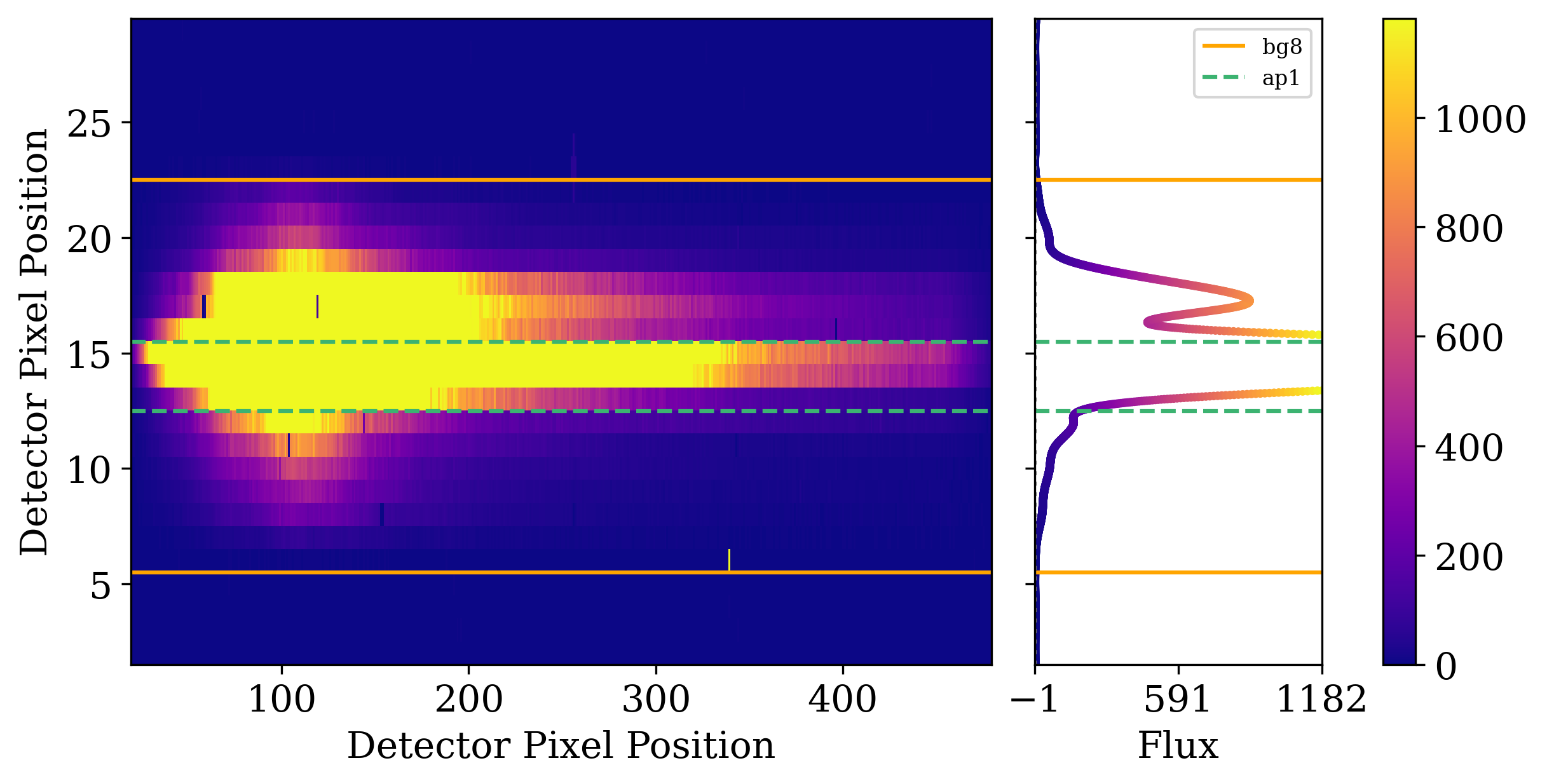}
    \caption{{\bf Imaged spectra.} Dispersed image (left) and vertical cut (right) of NGTS-10A and B.  
    The dashed green lines bound the 3-pixel aperture size and the solid orange lines indicate the inner edges of the background region (used in Stage 1). In order to minimize the contribution from NGTS-10B, we chose not to center the aperture on the peak contribution from NGTS-10A.  
    As seen in the right panel, a small amount of light from the binary star overlaps with the primary.
    }
    \label{fig:3304}
\end{figure}

{ Because the higher spatial resolution of our JWST observations, we detected that NGTS-10 was composed of two stars, NGTS-10A and NGTS-10B. Previous estimates of stellar parameters were based on the spectrum from the combined stars~\citep{McCormac2020}. In the following section, we use the JWST spectra of each star to refine both the stellar parameters.}
We performed double-PSF fitting on a per-column basis in order to extract individual spectra from each star (labeled Schlawin).  In parallel, we forwent masking the binary, increased the aperture full width to 13 pixels, and used Stage 4cal of the \texttt{Eureka!} pipeline to extract the combined calibrated stellar spectrum of NGTS-10A+B during secondary eclipse (labeled Stevenson). In comparing the two methods, the shape of the combined spectrum is the same; however, there is a noticeable, unexplained flux offset.  This illustrates the challenges in obtaining a reliable, absolute spectrum using JWST time-series data, especially in the case of blended multiple-star systems.
{ We note that the GAIA DR3 data cannot be used to estimate the system distance. That is due to the binary nature of the system, making the parallax measurement from GAIA unreliable. It is evident from the excess noise metric, which is 1.8e3 for our target, well above the proposed threesold of 2\citep{Gaia2016, Gaia2021}. } 

In the double-PSF fitting, we started with an automatically-produced \texttt{\_cal.fits} file from segment 2 from MAST using JWST version 1.12.5, SDP\_VER 2023\_2a, CRDS\_VER and 11.17.9 CRDS\_CTX \texttt{jwst\_1170.pmap}.
We used \texttt{webbpsf} \citep{perrin2014} version 1.4.0.
We calculate the PSF for the NIRspec instrument from 0.65~\micron\ to 5.15~\micron\ at a spacing of 0.05~\micron\ with an oversampling of 7 times the pixel sampling and shift and add 21 over-sampled pixels and re-normalize to generate a simulated dispersed PSF.
We then use a \texttt{scipy} \texttt{least\_squares} fitter to minimize the chi-squared of two blended PSFs for each 0.05~\micron\ wavelength column sample.
Finally, we save the amplitude of the PSF-multiplier for NGTS-10A and NGTS-10B needed to fit the spectrum as the calibrated flux.

We first fitted the two individual stellar spectra (from Schlawin) using a grid of Phoenix models. Given the low spectral resolution of our data, we are insensitive to the gravity and the metallicity of the star. Furthermore, there is a well-known degeneracy between distance and radius of the star during the fit. In order to reduce the explored parameter space, we fixed the metallicity to solar and used the mass-radius-T$_{eff}$ relationships from~\citep{Boyajian2012}, equations 8 and 10. Therefore we only fit for the distance and the effective temperatures of the two stars. We assume an absolute error corresponding to { $10\%$} of the flux. Our best fit leads to NGTS-10A being a K star with an effective temperature of 4436 K and NGTS-10B being a M star with an effective temperature of 3495 K (See Table~\ref{table:1}).  We also tried fitting the two stars with different distances, but the fit did not improve. We therefore believe that the two stars form part of a single system. 

We then fit the combined stellar spectrum (from Stevenson). We used the best fit results from the previous steps as priors to guide the joint fit, allowing the parameters to vary by up to $5\%$ compared to the previous values. As shown in Figure~\ref{fig:stellar-fit}, we find a satisfactory fit to the Stevenson case. We note that the models are least consistent with the data around the peak of the SED. This could
be due to insufficient non-linearity correction in the data or inadequate stellar models with missing physics. We verified that this mismatch was not affecting significantly our planetary spectrum.  While both methods yield similar results, we adopt as final parameters those listed in the Stevenson column of Table~\ref{table:1}. These parameters correspond to a slightly hotter star than in~\citep{McCormac2020} ($4567K$ vs. $4400K$). 

\begin{figure}[h!]
    \centering
    \includegraphics[width=0.97\linewidth]{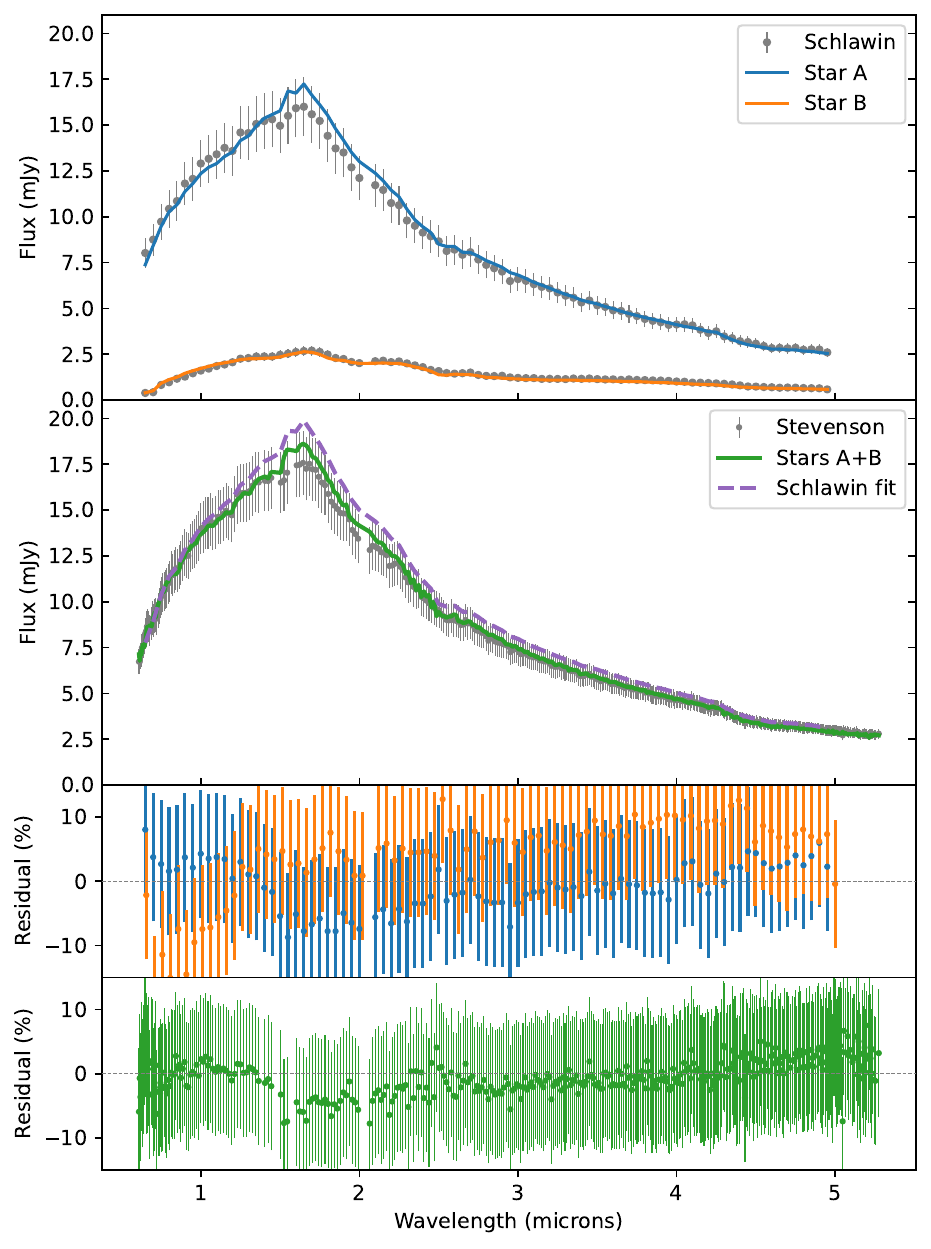}
    \caption{{\bf Stellar spectrum.}Stellar spectrum from the Schlawin reduction (first panel) and the Stevenson reduction (second panel) together with the best fit Phoenix models described in Table~\ref{table:1}.{ We assume errorbars on the absolute flux of $10\%$} The corresponding residuals are shown in the bottom two panels.}
    \label{fig:stellar-fit}
\end{figure}

\begin{table}
\caption{Stellar parameters}             % title of Table
\label{table:1}      % is used to refer this table in the text
\centering                          % used for centering table
\begin{tabular}{l| c c}        % centered columns (4 columns)
\hline                 % inserts double horizontal lines
 & Schlawin & Stevenson \\    % table heading 
\hline \hline                       % inserts single horizontal line
 $T_{\rm A} (K)$ & $4436\pm50$ & $4567\pm36$  \\      % inserting body of the table
 $R_{\rm A}^{*}$ $(R_{\rm \odot})$ & $0.68\pm0.03$ & $0.70\pm0.03$\\      % inserting body of the table
$T_{\rm B}(K)$ & $3495\pm44$ & $3670\pm146$  \\
 $R_{\rm B}^{*}$ $(R_{\rm \odot})$ & $0.39\pm0.03$ & $0.40\pm0.03$\\      % inserting body of the table
distance (PC) & $305\pm50$ & $346\pm18$     \\
\hline                                   %inserts single line
\end{tabular}
$^{*}$ The mean value is calculated using eq. 8 of~\citep{Boyajian2012} and the error is estimated based on the spread of the population around the relationship. 

\end{table}

We note that our extracted stellar spectra still contains $3.6\%$ of the dayside planetary spectra due to the grazing geometry of the system. Given that the planet-to-star flux ratio is, at most, $0.74\%$, the additional planetary contribution corresponds no more than $0.027\%$ of the stellar spectra and is unlikely to affect our determination of the stellar parameters. 

\subsection*{Light Curve Fitting} \label{sec:fitting}

{ The band-integrated (white) light phase curve spans 0.6 to 5.4 $\mu$m.  We explored both fixed and free quadratic limb darkening parameters, the former using ExoTiC-LD \citep{Grant2024} and the Stagger 3D model grids \citep{Magic2015}. Due to the planet's grazing nature, the limb darkening parameters are difficult to constrain using the NIRSpec data.  The final results shown in \ref{tab:white_parameters} assume fixed parameter values. When generating our 47 spectroscopic phase curves, we manually mask several pixel columns (117, 125, 339, 354, 385, and 447) that exhibit excessive noise.  The corresponding wavelengths of the masked columns are: 1.426, 1.587, 4.387, 4.521, 4.784, and 5.268 $\mu$m.}  The data span 0.6 to 5.3 $\mu$m, which corresponds to 100-nm wide bins.

In the lightcurve fitting stage, we mask the first 1,000 integrations ($\sim$27\,minutes) from the start of each exposure (e.g., see \ref{fig:wlc}) and fit for a wavelength-dependent offset between exposures.  We apply BATMAN \citep{Kreidberg2015b} to fit both the transit and eclipse, a four-parameter phase curve model to allow for asymmetric phase variations, a linear trend in time to account for time-dependent systematics, { two flux offsets between exposures, and a scatter multiplier to account for excess noise above theoretical predictions.  The phase variation component takes the functional form:
\[P(\phi) = 1 + A_1(cos(\phi)-1) + A_2sin(\phi) +
A_3(cos(2\phi)-1) + A_4sin(2\phi)
\]
where $\phi$ is the planet's orbital phase and $A_N$ are the free parameters representing the amplitude of each sinusoid. }For the white light phase curve, we have 15 free parameters.  We list the fixed, best-fit, and derived astrophysical parameters from the white light curve fit in \ref{tab:white_parameters}. We estimate the planet radius by combining our measured radius ratio and the stellar radius from \ref{table:1} (Stevenson column).  After fixing the transit midpoint, eclipse midpoint, inclination, and semi-major axis, each spectroscopic phase curve has 11 free parameters.  We use EMCEE to determine the uncertainty on each free parameter.

\begin{table}
    \centering
    \caption{Fixed, best-fit, and derived astrophysical parameters from our white (0.6 - 5.4 $\mu$m) light curve.}
    \begin{tabular}{ll|c}
        \hline
        % FIXED LD SOLUTION (0.6 - 5.4 microns)
        Fixed &
        Period (Days)             & 0.7668944     \\
        & Eccentricity            & 0             \\
        & $u_1$                    & 0.22933     \\
        & $u_2$                   & 0.22831     \\
        \hline
        Best-Fit &
        Transit Time              & 2460015.082427(13)\\
        & (BJD$_{TDB}$)           & \\
        & Eclipse Time            & 2460014.69938(18) \\
        & (BJD$_{TDB}$)           & \\
        & Radius Ratio            & $0.1819\pm0.0007$ \\ 
        & Inclination ($\degree$) & $79.40 \pm 0.05$  \\
        & $a/R_s$                 & $4.587 \pm 0.011$ \\
        & Eclipse Depth (\%)      & $0.153 \pm 0.003$ \\
        & $A_1$                   & $0.470 \pm 0.015$ \\
        & $A_2$                   & $-0.111 \pm 0.007$  \\
        & $A_3$                   & $0.011 \pm 0.007$  \\
        & $A_4$                   & $0.019 \pm 0.006$  \\
        \hline
        Derived &
        Transit Depth (\%)        & $3.31 \pm 0.03$ \\
        & $R_{\rm p}$ $(R_{\rm J})$ & $1.24 \pm 0.05$ \\
        &Equilibrium temperature* (K)  & $1508\pm12$\\

        % FREE LD SOLUTION
        % Fixed &
        % Period (Days)             & 0.7668944     \\
        % & Eccentricity            & 0             \\
        % \hline
        % Best-Fit &
        % Transit Time              & 2460015.082427(13)\\
        % & (BJD$_{TDB}$)           & \\
        % & Eclipse Time            & 2460014.69938(19) \\
        % & (BJD$_{TDB}$)           & \\
        % & Radius Ratio            & $0.1820\pm0.0010$ \\ 
        % & Inclination ($\degree$) & $79.43 \pm 0.14$  \\
        % & $a/R_s$                 & $4.592 \pm 0.026$ \\
        % & $u_1$                   & $0.27^{+0.14}_{-0.17}$\\
        % & $u_2$                   & $0.19^{+0.18}_{-0.14}$      \\
        % & Eclipse Depth (\%)      & $0.152 \pm 0.004$ \\
        % & $A_1$                   & $0.471 \pm 0.017$ \\
        % & $A_2$                   & $-0.112 \pm 0.008$  \\
        % & $A_3$                   & $0.012 \pm 0.007$  \\
        % & $A_4$                   & $0.019 \pm 0.006$  \\
        % \hline
        % Derived &
        % Transit Depth (\%)        & $3.31 \pm 0.04$ \\
        % & $R_{\rm p}$ $(R_{\rm J})$ & $1.23 \pm 0.06$ \\
        % % & $R_{\rm p}$ $(R_{\rm J})$ & $1.237^{+0.007}_{-0.003}$\\
        \hline
    \end{tabular}
            *For full redistribution and zero albedo

    \label{tab:white_parameters}
\end{table}

{ In order to achieve a reduced chi-squared of unity, we apply the wavelength-dependent multiplicative factors shown in \ref{fig:scatter_mult} to the flux uncertainties inferred from our data reduction step.  Values range from 1.2 -- 1.8, with larger multiplicative factors at shorter wavelengths.
\ref{fig:CorrNoise} depicts the level of correlated noise in the spectroscopic residuals.  We find that the residuals are dominated by white noise at $>1.7$ {\microns}.  At shorter wavelengths, red noise appears to have a non-negligible effect on the residuals.
}

\begin{figure}[ht]
    \centering
    \includegraphics[width=0.9\linewidth]{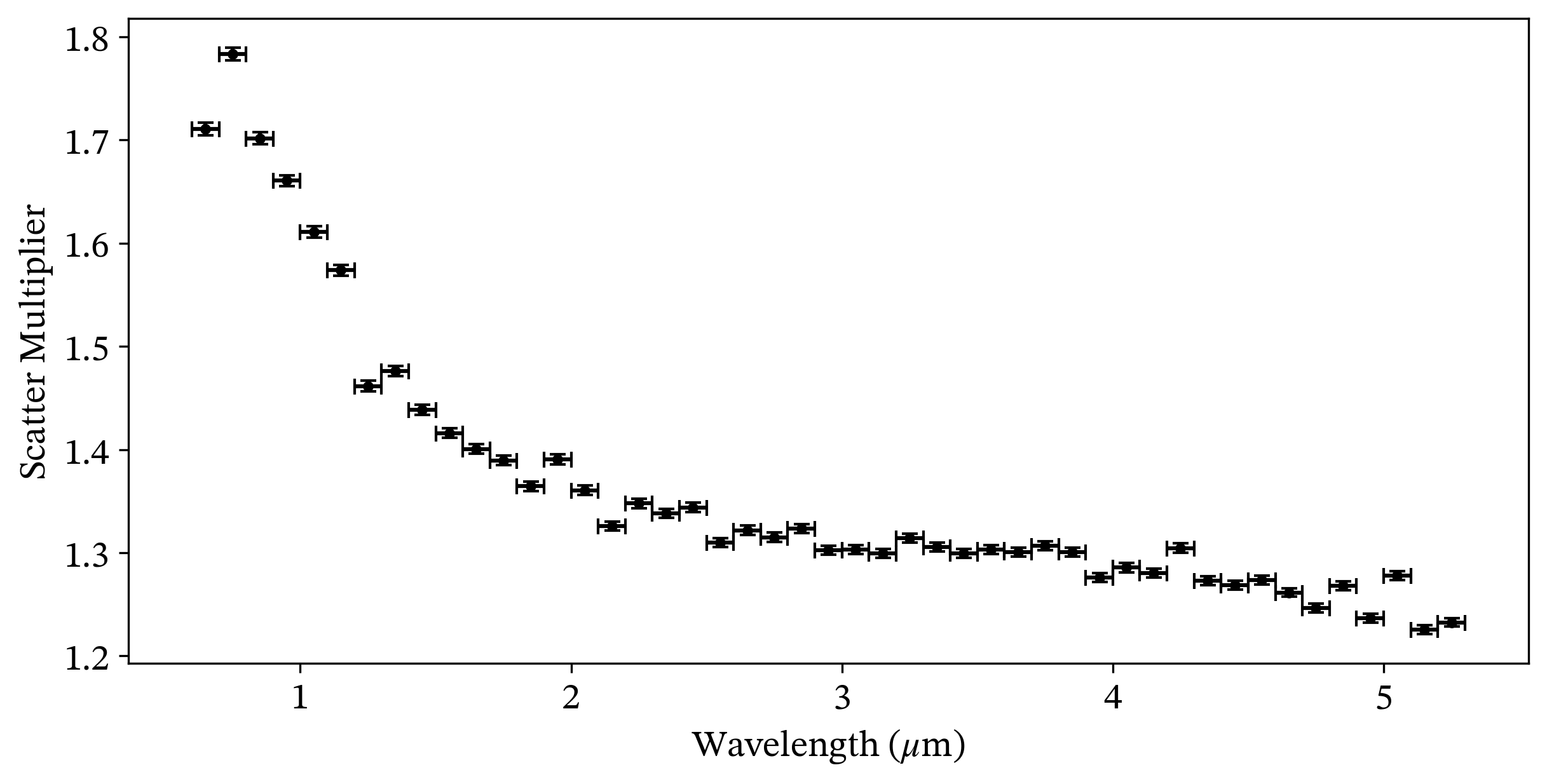}
    \caption{Uncertainty enhancement relative to the computed noise.  As part of the spectroscopic light curve fitting step, we adopt the scatter multiplier as a free parameter. The errobars represent the bin width.}
    \label{fig:scatter_mult}
\end{figure}

\begin{figure}[ht]
    \centering
    \includegraphics[width=0.9\linewidth]{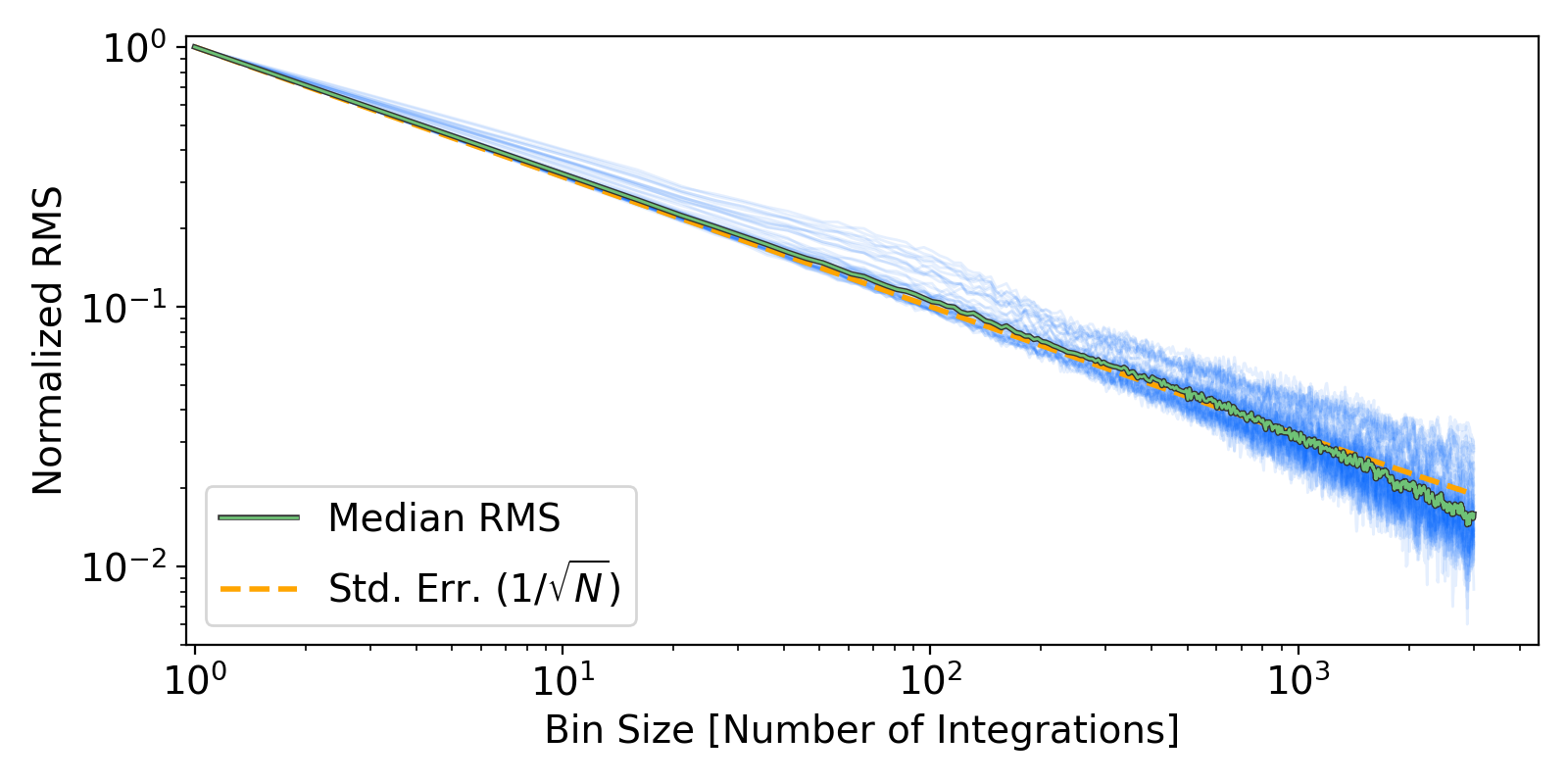}
    \caption{{\bf Observed correlated noise.}Correlated noise versus bin size from the spectroscopic light curve fits.  The blue curves depict the level of time-dependent correlated noise in the residuals from all 47 spectroscopic channels.  The median RMS (solid green curve) closely follows the standard error expectation (dashed orange line), thus indicating no significant red noise for many of the channels. There is, however, evidence for red noise in the 11 spectroscopic channels $<1.7$~{\microns}.}
    \label{fig:CorrNoise}
\end{figure}

\subsection*{Interior model}
The intrinsic temperature $T_{\rm int}$ represents the total radiative cooling luminosity of a planet's interior. We construct a one-dimensional interior model to evaluate possible values of $T_{\rm int}$.  The model assumes a pure H/He envelope atop an Earth-like rock/iron core, consisting of a distinct silicate mantle and a metallic inner core \citep{Tang2025}. The model solves for hydrostatic equilibrium, with an adiabatic H/He envelope, and with the core temperature structure calculated via parameterized convection.  The radiative atmosphere above the interior is assumed to have solar metallicity, with $T_{\rm int}$ determined following a modeling grid (\citep{Fortney2007}) that connects the envelope's specific entropy and gravity to its $T_{\rm int}$. The atmosphere's temperature-pressure-radius profile is evaluated with an analytical double-stream gray atmosphere model \citep{Guillot2004}, which self-consistently sets the radiative-convective boundary. The planetary radius is defined at the 20\,mbar pressure level, typical for optical transits in a dust-free atmosphere \citep{Lopez2012}. 

{ To constrain the values of $T_{\rm int}$, we calculated the radius of NGTS-10b for various values of internal temperature and core mass. As shown in Figure~\ref{fig::interior}, the radius of the planet increases as the internal temperature increases and decreases as the core mass of the planet increases. Based on a single radius measurement, the two parameters are therefore degenerate. We find that for a $20M_{\rm Earth}$ core, often predicted by core-formation models, the solutions compatible with our radius measurement have an internal temperature ranging from $250$ to $400K$. For a unexpectedely large core mass of $60 M_{\rm Earth}$, the internal temperature can reach, at maximum, $500K$. Thus, we find that $T_{\rm int}$ values above 500\,K are unlikely.}

\begin{figure}
\includegraphics[width=0.7\linewidth]{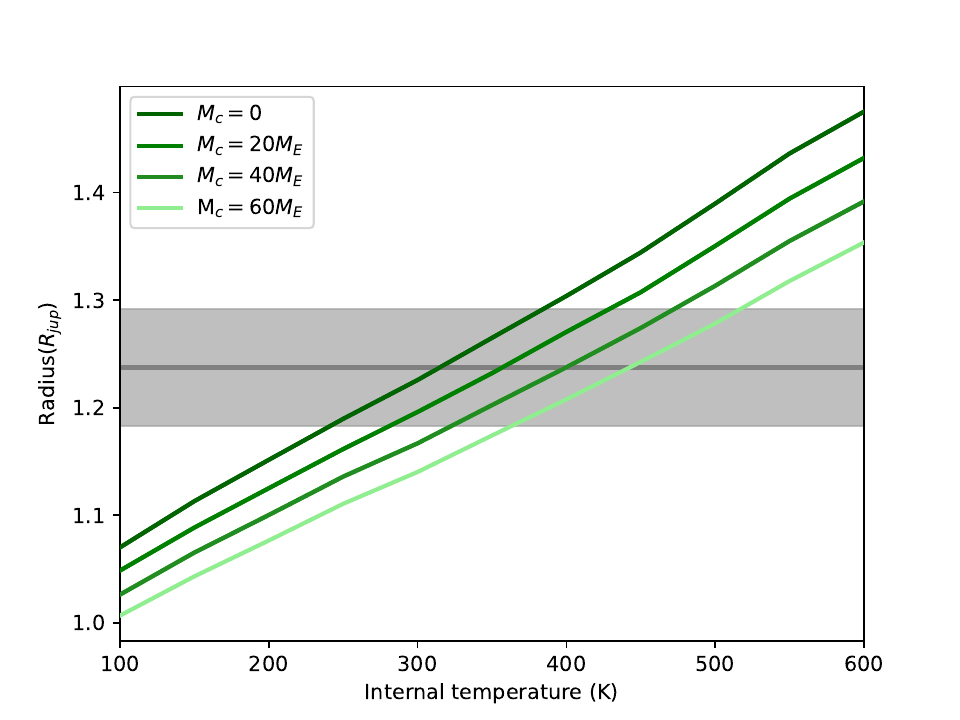}
\centering
\caption{{\bf NGTS-10A b interior modeling.} NGTS-10b radius as a function of it's internal entropy, parameterized as the internal temperature, calculated for different core mass. The gray line and gray zone represents the mean and the $1\sigma$ confidence interval on the measured planetary radius. Internal temperatures higher than $\approx 500K$ would need a much larger core than predicted by theory to fit the planet radius and are thus unlikely. 
}
\label{fig::interior}
\end{figure}
\subsection*{Atmospheric Models} \label{sec:retrievals}

\paragraph{Radiative/convective models}\label{sec:LuisRetrieval}
We use the radiative/convective model ScCHIMERA to model the emission spectrum of NGTS-10A b. The model solves for the thermal structure of the atmosphere within the assumption of radiative, convective, and chemical equilibrium. The model is iterated from an initial solution using a Newton-Raphson method until convergence is reached. The model has been used extensively to simulate the thermal structure and spectra of hot Jupiters \citep{Piskorz2018, Bell2023, Wiser2026, Welbanks2024}.

We performed a grid of models varying irradiation, C/O, and metallicity and { used a nested sampling algorithm over the interpolated grid} to estimate the best fit parameters. The grid additionally parameterized clouds through a constant, grey opacity. 

Our best fit of the dayside spectrum has $[\text{M/H}]=-0.05^{+0.04}_{-0.07}$ and $\text{C/O}=0.39^{+0.05}_{-0.05}$. The best fit dayside { effective} temperature is $1797^{+2}_{-4}$~K. This is smaller than the $1829$~K obtained from the observations directly. That is because the grid retrieval aims at finding the best fit of the spectra and does not enforce the conservation of the total emitted flux. The fit is therefore biased to fit the longer wavelengths where the errorbars are smaller. Overall, the radiative/convective model is a good fit to the longer wavelengths, but for the shorter than 2 {\microns} range the data does not show the large water absorption bands present in the model. This is puzzling as the contribution functions for the water bands at 1.5 and 1.8 microns overlap with the contribution function of the water band between 2 and 3 {\microns}. We tested whether the presence of an additional grey absorber limited to the region below $2\mu m$ could reduce the water bands and provide a better fit to the data. However, such model, in radiative/convective equilibrium, leads to a near-isothermal atmosphere (as predicted for semi-grey atmospheres~\citep{Guillot2010}) and thus reduces significantly the water feature strength at longer wavelengths. We therefore postulate that the apparent mismatch at short wavelengths is due to an increased scatter of the data points in that region caused by the smaller signal-to-noise ratio. Overall, our determination of the chemical composition of the atmosphere relies mostly on the wavelengths longer than 2 {\microns} and are thus not too affected by the mismatch in the shorter than 2 {\microns} range.

\paragraph{Free Retrieval}\label{sec:LuisRetrieval}

We further infer the atmospheric properties of NGTS-10A b using parametric models without restrictions imposed by radiative-convective-thermo-chemical-equilibrium as with the ScCHIMERA modeling described above. This modeling paradigm, also known as ``free-retrievals'' allows us to infer the chemical composition of the planet and the vertical temperature structure of the planet and obtain a goodness-of-fit metric to perform model comparisons (see ref~\citep{Welbanks2023}).  We use the retrieval framework Aurora\citep{Welbanks2021}, a code for the interpretation of transmission \citep{Welbanks2022, Welbanks2024} and emission spectra \citep{Bell2023} of exoplanets.

The models used to interpret the nightside and dayside spectra assume line-by-line opacity sampling and are computed at a spectral resolution of R=10,000. The chemical absorbers considered are H$_2$O\citep{Rothman2010}, CH$_4$\citep{Yurchenko2014}, NH$_3$\citep{Yurchenko2011}, HCN\citep{Barber2014}, CO\citep{Rothman2010}, CO$_2$\citep{Rothman2010}, and SO$_2$\citep{Underwood2016}. The models do consider opacity due to H$_2$-H$_2$ and H$_2$-He collision induced absorption\citep{Richard2012}. The atmospheric extent in the models is from 100\,bar to 10$^{-6}$\,bar, using 50 discretized layers. The vertical pressure-temperature structure is parameterized following the 6-parameter prescription from ref.\citep{Madhusudhan2009}. The Bayesian parameter estimation is performed using PyMultiNest\citep{Buchner2014}. The priors for the chemical abundances are uniformly distributed in log-space between volume mixing ratios of $10^{-12}$ and $10^{-1}$ following ref~\citep{Welbanks2019} and imposing a physical restriction where the CO abundance has to be smaller than the H$_2$O abundance, consistent with a C/O ratio lower than 1, as in ref~\citep{Welbanks2021}. For the temperatures of NGTS-10b, the abundance of CO is always predicted to be lower than that of H$_2$O regardles of C/O ratio \citep{Madhusudhan2012}, and metallicity \citep{moses_compositional_2013} under chemical-equilibrium expectations. The temperature profile priors are as described in previous works \citep{Welbanks2019,Bell2023} with the exception of T$_0$, the temperature at the top of the model atmosphere, which has a uniform prior from 300~K to 2000~K.

%%%%%%%%%%%%%%%%%%%%%%%%%%%%%%%%%%%%%%%%%%%%%%%%%%%%%%%%%%%%%%%%%%%%%%%

\subsubsection*{Data Availability}
The data used in this paper are associated with JWST GTO program 1185 (PI Parmentier) and is publicly available from the Mikulski Archive for Space Telescopes (\url{https://doi.org/10.17909/fa95-k770}). The reduced data used to produce Fig. 2 is available on Zenodo (\url{https://doi.org/10.5281/zenodo.18861579}).

\end{document}